\documentclass{egpubl}
\usepackage{eurovis2024}

\EuroVisShort  

\usepackage[T1]{fontenc}
\usepackage{dfadobe}

\usepackage{cite}  
\BibtexOrBiblatex
\electronicVersion
\PrintedOrElectronic
\ifpdf \usepackage[pdftex]{graphicx} \pdfcompresslevel=9
\else \usepackage[dvips]{graphicx} \fi

\usepackage{egweblnk}

\begin{document}

\title{Optimizing Mentor-Student Communication with Symbolic Design for Message States}
\author[D. Fellner \& S. Behnke]
{\parbox{\textwidth}{\centering 
Yuanzhe Jin$^{1}$ and Jiali Yu$^{2}$ 
        }
        \\
{\parbox{\textwidth}{\centering 
$^1$ University of Oxford, Oxford, England, United Kingdom\\
$^2$ Zhejiang University, Hangzhou, Zhejiang, China
       }
}
}

\maketitle

\begin{abstract}
In the mentor-student communication process, students often struggle to receive prompt and clear guidance from their mentors, making it challenging to determine their next steps. When mentors don't respond promptly, it can lead to student confusion, as they may be uncertain whether their message has been acknowledged without resulting action. Instead of the binary options of "read" and "unread," there's a pressing need for more nuanced descriptions of message states. To tackle this ambiguity, we've developed a set of symbols to precisely represent the cognitive states associated with messages in transit. Through experimentation, this design not only assists mentors and students in effectively labeling their responses but also mitigates unnecessary misunderstandings. By utilizing symbols for accurate information and understanding state marking, we've enhanced communication efficiency between mentors and students, thereby improving the quality and efficacy of communication in mentor-student relationships.


\end{abstract}

\keywords{\textbf{Mentor-student Relationship}, \textbf{Human-computer Interaction}, \textbf{State Design}}

\maketitle
\section{Introduction}
Effective communication plays a vital role in fostering productive mentor-student relationships. However, in the context of interactions between mentors and students, challenges often arise in ensuring timely and clear instructions for the next steps. Regardless of the communication mode employed, whether it be email, text messaging, phone calls, or video conferences, students frequently encounter difficulties in receiving prompt and explicit guidance from their mentors. This may manifest as unanswered or unread messages, leaving students uncertain about their mentors' level of understanding or whether any concrete actions have been taken to address their questions or concerns. As the foundation of mentorship is built upon the pillars of guidance and support, these communication gaps can impede the overall effectiveness of the mentoring process. In the fast-paced and dynamic world of academia and professional development, such uncertainties can be detrimental to a student's progress and growth. 

Recognizing the importance of addressing these communication issues and enhancing the mentor-student relationship, this research introduces a set of symbols meticulously designed to provide a more accurate representation of the cognitive states associated with receiving and processing information. By employing these symbols, students and mentors alike can better label the status of their information exchanges, thus reducing the likelihood of misunderstandings and optimizing the efficiency of their interactions. This solution aims to bridge the gap between immediate responses and no responses at all, offering a precise description of information flow during mentor-student communication.

Through empirical experiments, this design not only aids in improved information state recognition between mentors and students but also mitigates the unnecessary ambiguities that often plague such interactions. Students will gain greater clarity on their mentor's level of understanding and engagement. Meanwhile, mentors can more accurately signal the status of guidance requested by students. By employing these symbols to achieve a more precise shared understanding of information states, the communication efficiency between mentors and students can be significantly enhanced. Response lags can be reduced, and clarity improved.

The idea of the design has the ability to elevate the quality and effectiveness of mentorship across academic disciplines and professional development contexts. With enhanced communication, students gain the timely guidance needed to progress in their learning journeys. Mentors can provide support more efficiently to a broader range of students. The research contributes an innovative solution that moves beyond the limitations of existing communication tools, bridging gaps and creating shared meaning to optimize mentor-student relationships. This has far-reaching implications, as high-quality mentoring plays a key role in student success, knowledge transfer, and the cultivation of talent in many fields. By improving communication, this work lays the foundation for more productive mentorships and better outcomes.
\section{Related Work}
\subsection{Factors Affecting Mentor-Student Relationships}

Studies have shown that the discipline category is an important structural condition that influences the mentor-student relationship \cite{Austin:2002:education, Golde:2006:envisioning}, as the feature of disciplines’ knowledge production patterns leads to differences in the daily research interaction behaviors of mentors and their students. For example, unlike mentors and students who majored in the natural sciences, who often work collectively in laboratories and have immediate discussions, students who majored in the social sciences tend to receive one-on-one guidance in the mentor’s office \cite{Noy:2012:graduate}, where interviews and discussions are more formalized. The gender factor has also received academic attention \cite{Mutchnick:1991:mentoring}, with some researchers empirically concluding that mentoring between mentors and students of the same gender is more effective than the opposite. Some scholars have also examined the impact of specific details of the mentoring model, such as the frequency of mentoring \cite{Kumar:2017:mentoring}, whether face-to-face \cite{Nasiri:2015:postgraduate}, and the content of support. Frequent interactions help increase trust between mentors and students \cite{De:2013:role}. Online mentoring faces challenges such as technical difficulties, time management, difficulty writing and receiving written feedback, and life events interrupting study \cite{Pollard:2021:mentoring}. The studies shed light on our efforts to clarify the communication problems encountered by mentors and students in specific research scenarios. In addition, distinctions of mentor support behaviors and functions provide multiple perspectives for us to examine the roles played by symbols in communication, e.g., psychosocial support helps to increase student satisfaction with their mentors, while instrumental support helps to improve the efficiency of research work \cite{Tenenbaum:2001:mentoring}.

Mentoring relationship research is based on empirical methods, and the most commonly used are semi-structured interviews, participant observation, and questionnaire methods. Most of the studies using the first two methods point to qualitative analysis. The researcher lists the outline of the interview and disperses, focuses, and refines the questions in the survey process \cite{Lechuga:2011:faculty}. Through participatory observation in real-life scenarios, to grasp the details of the interactions of the interviewees’ facial expression, Posture, and tone of voice \cite{Mehrabian:2017:communication}, picking a particular university or several universities as a sample frame, doing questionnaire design and data analysis with the help of tools such as scales, SPSS and so on \cite{Blaney:2020:mentoring}.


\subsection{Symbol-Based Information State Design}

In the design of tactile symbols, a recent study \cite{Sabnis:2023:tactile} explored the potential of using continuous and motion-coupled vibrations. The study showed that personal experiences and emotions significantly influence symbol design. This research emphasized the importance of designing tactile symbols based on usage contexts. In data visualization, a research group \cite{Jin:2023:radial} proposed the concept of radial icicle trees that maintain consistent node areas while enhancing the visibility of narrow rectangles. This provides new insights for displaying data with inherent connections. Regarding differentiating data sizes, a team of researchers \cite{Li:2010:model} studied factors affecting size perception in scatterplots and found approximate homogeneity in complex tasks. Another study \cite{Tohidi:2006:ACM} analyzed usability testing of a single interface with three designs that had the same functionality but different styles. These studies provided foundations for designing better tactile and visualization symbols. 

In practical applications, a group of researchers \cite{Zhao:2018:coding} designed tactile symbols representing phonemes and examined human capabilities in learning and processing tactile information. Horton et al. \cite{Horton:1996:designing} discussed guidelines for clear and comprehensible icon design. A recent study \cite{Haoran:2023:magical} combined symbolic culture with AI generation to create modern Chinese paintings. These works applied symbolic design to real-world scenarios. Another research team \cite{Roque:2009:ACL} explored the use of information states in call-for-fire situations.

Regarding the integration of symbols and learning, a study \cite{Pinhanez:2021:integrating} examined combining expert knowledge with machine learning training to improve dialogue system accuracy. A group of researchers \cite{Vaataja:2016:information} proposed information visualization design guidelines tailored for expert evaluation. These two studies provided examples of integrating symbolic knowledge and learning methods.

In dialogue systems, a research team \cite{Devault:2005:information} constructed a modular architecture incorporating information states. Another study \cite{Gavsic:2011:effective} modeled dialogues as Markov decision processes, emphasizing the importance of logging user goals and histories. These two studies laid the foundations for dialogue management and modeling. In information retrieval, a group of researchers \cite{Liu:2021:state} identified user task states and conducted adaptive evaluation accordingly. This highlighted the significance of considering user states in retrieval. Another research team \cite{Varges:2008:DD} proposed a new data-centric dialog system and conducted user research at the university's front desk.

Current research has extensively explored symbolic usage and design in aspects including symbol design, application, integration with learning, dialogue modeling, etc., establishing foundations in this field. However, further studies on the in-depth integration of symbols with learning, interaction, etc. are still needed to uncover their potential.

\section{Background and Designs}

\subsection{Preliminary Research and User Interviews}

We conducted interviews with several current students to obtain initial insights into the communication methods employed by this cohort when interacting with their mentors regarding their academic work. The majority of mentors preferred using instant messaging tools or face-to-face meetings to discuss work-related tasks such as paper revisions and project progress. Some mentors also used email as a supplementary means for transmitting, receiving, and storing documents.

The students we interviewed acknowledged the advantages of instant messenger in communication. They noted that these tools enhanced the efficiency of their research work and fostered a more consistent emotional connection with their mentors. For instance, Participant A pointed out that minor research queries could often be quickly resolved through brief exchanges on instant messenger, demonstrating its high effectiveness. Participant B mentioned that their mentor would send encouraging messages via instant messenger, conveying care and support.

However, interviewees also highlighted limitations associated with instant messaging. For example, Participant C expressed the view that instant messenger was primarily suitable for short, informal conversations and less suitable for in-depth research discussions that require rigor. Additionally, Participant D shared a personal experience wherein his mentor occasionally went offline on instant messenger, leaving him uncertain whether to await a reply or return to his research work. Participant D also mentioned situations where the mentor had promised to schedule a follow-up discussion on instant messenger but later struggled to keep track of this commitment due to a growing workload, causing delays in the research progress.

\subsection{Information States}

\begin{figure}[htbp]
  \centering
  \includegraphics[scale=0.23]{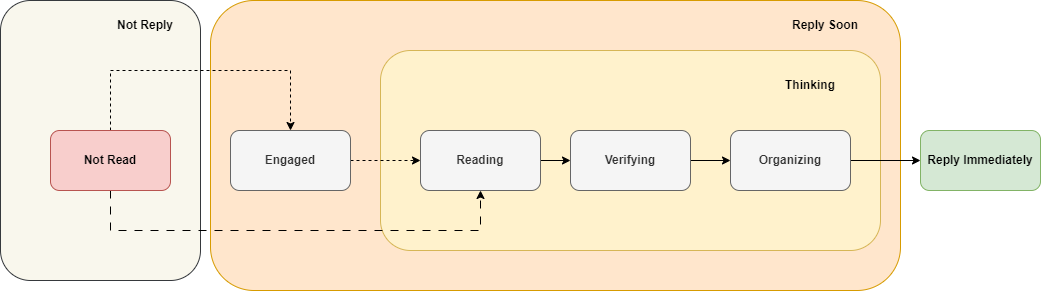}
  \caption{Information states transition}
\label{workflow}
\end{figure}

The insights gleaned from our interviews with graduate students and their mentors shed light on the complexities surrounding communication within these academic relationships. Existing instant messaging tools available in the market, such as Slack, WhatsApp, Messenger, and others, typically employ the use of emoticons (emoji) to indicate the status of messages \cite{Arafah:2019:language}. However, through in-depth discussions with students and their mentors, we have realized that the simple use of markers like "not read" and "checked" presents challenges in accurately understanding the intent behind a message. Between these two states, there exist more intricate and precise information states that can better convey a mentor's expectations and intentions regarding the message.

To gain a deeper understanding of the complexity of information states, we drew inspiration from the steps humans undergo in information processing. We proposed an information states transition diagram in Fig. \ref{workflow} that elaborately illustrates the process of information reply starting from the "not read" state, further breaking it down into multiple specific states. These states include the initial "Engaged" state, signifying that the mentor is currently occupied with other tasks, moving to the state of having seen the message but not being able to reply immediately, and then gradually transitioning through the states of "Reading," "Verifying," and "Organizing" thoughts, ultimately leading to the action of a "Reply Immediately". It is worth noting that even in situations where an immediate reply is chosen, these four states still sequentially occur, reflecting the steps the human brain undergoes in processing information and achieving the final response.

This multi-stage understanding of information states helps us delineate the information exchange states between mentors and students more accurately, enabling us to better comprehend the intricacies of information processing and reply in the communication process. By using symbols to represent these states in information exchange, we significantly enhance the clarity and efficiency of communication, while reducing the risk of information misinterpretation, thereby optimizing the quality and effectiveness of mentorship relationships.




\subsection{Symbolic Designs}

\begin{figure}[htbp]
  \centering
  \includegraphics[scale=0.15]{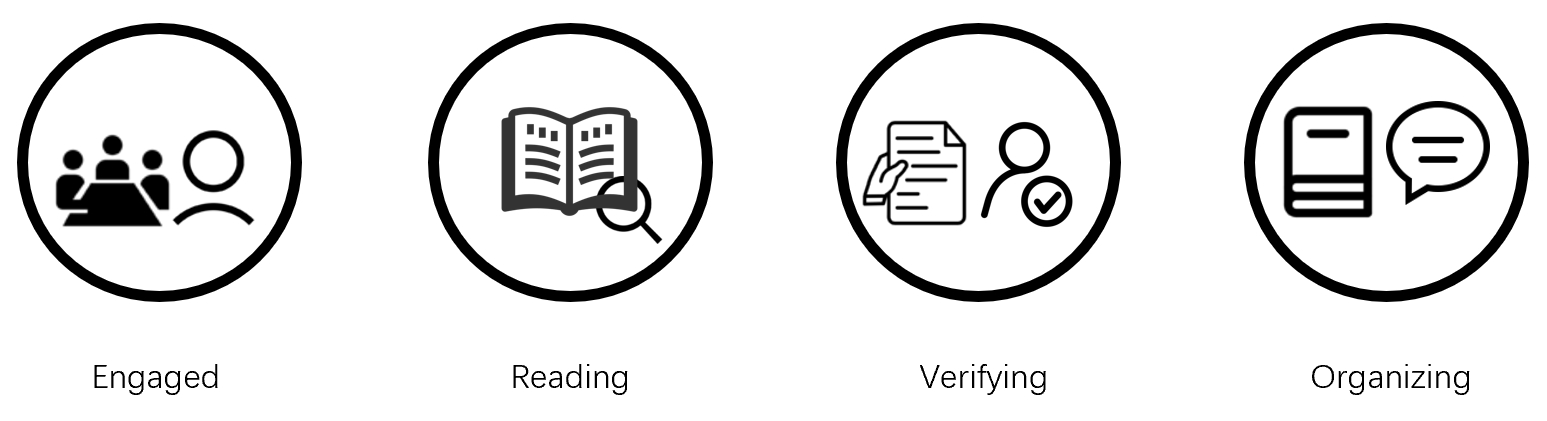}
  \caption{A symbol set for describing the four message states}
\label{symbol}
\end{figure}

In response to the four information states, we have designed specific symbols to depict the different states of information exchange between mentors and students. These symbols have been meticulously crafted, taking into consideration the distinctive characteristics of each information state to ensure they convey the cognitive status of the information recipient. Specifically, we have designed the following four symbols, which can be observed in Fig. \ref{symbol}:
\begin{enumerate}
    \item \textbf{Engaged} Symbol: This symbol features a person alongside a group engaged in a meeting, clearly conveying that the recipient is currently occupied with other tasks and unable to reply immediately. This is crucial for avoiding interruptions to the recipient's ongoing work.

    \item \textbf{Reading} Symbol: This symbol includes icons of a magnifying glass and a book, explicitly indicating that the information recipient is carefully reading the content of the message. This helps the sender understand that the recipient is attentively processing the information rather than ignoring or disregarding it.

    \item \textbf{Verifying} Symbol: Comprising a checklist and a confirmation symbol, this symbol signifies that the information recipient has received the message but requires further verification of its accuracy. This helps reduce the likelihood of making erroneous decisions based on incomplete or uncertain information.

    \item \textbf{Organizing} Symbol: Composed of a combination of a book and a text bubble, this symbol signifies that the recipient needs some time to organize their thoughts and replies. This helps prevent hasty replies and confusion.
\end{enumerate}

The use of these symbols serves multiple positive purposes in information exchange between mentors and students. Firstly, they provide clear and consistent markers for information states, aiding both parties in better understanding each other's cognitive status. This helps eliminate uncertainty and misunderstandings in information communication, enhancing the accurate conveyance of information.

Secondly, the design of these symbols contributes to improved communication efficiency. Recipients can use symbols to convey their current status, informing senders when to expect a reply. This reduces unnecessary waiting time, enabling both parties to make decisions and take action more promptly. These symbols foster mutual trust between mentors and students. Students know that mentors are diligently processing information, and mentors gain a better understanding of students' needs and concerns. Such trust is crucial for building robust mentor-student relationships and supporting students' academic and career development.

In the following sections, we delve deeper into the specific impacts of these symbols in two real-world application scenarios and their effects on mentorship relationships. Through user studies, we gain a better understanding of the way that symbol design improves information exchange and provides valuable guidance for future research and practice.

\section{User Study}

In the user study, we selected two common mentoring scenarios, namely the thesis revision guidance scenario and the literature review guidance scenario, as pivotal contexts for the research. This choice was based on the ubiquity and significance of these two scenarios in various academic disciplines and mentorship relationships. They represent typical activities within mentorship relationships, and these activities have widespread applicability across diverse academic fields. This implies that the research findings and symbol designs possess a generalizability that extends to different domains and mentorship relationships, offering practical value to a broader audience. By focusing on these key scenarios, we aim to provide targeted and feasible solutions for enhancing information exchange within mentorship relationships, ultimately contributing to the improvement of mentorship quality and effectiveness.


\subsection{User Background}

\begin{figure*}[htb]
  \centering
  \begin{tabular}{@{}c@{\hspace{4mm}}c@{\hspace{4mm}}c@{}}
    \includegraphics[width=43mm,height=30mm]{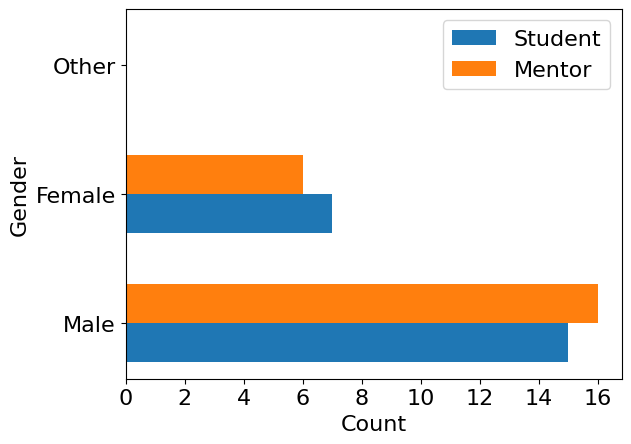} &
    \includegraphics[width=43mm, height=30mm]{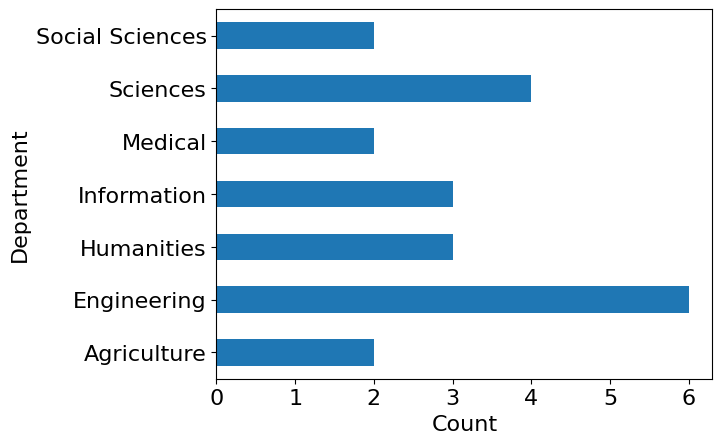} &
    \includegraphics[width=43mm,height=30mm]{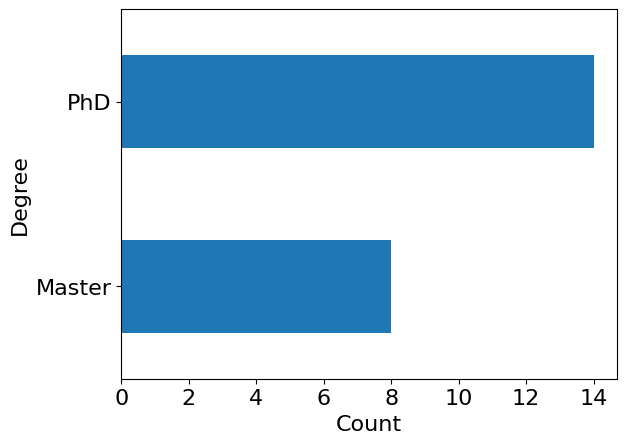}\\
    \small (a)  Mentor-Student Gender Distribution &
    \small (b)  Department Distribution &
    \small (c)  Degree Distribution
  \end{tabular}
  \caption{Overview of participant demographic information}
  \label{user background}
\end{figure*}

We had established our objective of investigating whether precise representations of information states contribute to more harmonious and efficient communication between mentors and students, we conducted a study involving 22 participants. To comprehensively assess the impact of these symbols, we sought a diverse range of participants, considering factors such as age, academic discipline, mentorship arrangements, and gender. The demographic information of our participants is detailed in Fig. \ref{user background}, demonstrating representation across seven different academic disciplines, spanning from the humanities and engineering to the natural sciences. Additionally, our participant pool encompassed a spectrum of educational backgrounds, ranging from postgraduate students to doctoral students.

To address this research question, we adopted a multifaceted approach involving both qualitative and quantitative methods. Initially, we administered our newly designed symbols to the participants and conducted exploratory interviews with them. These interviews aimed to gather their perspectives, emotions, and behaviors when presented with these symbols during communication. Building upon the insights gained from the interviews, we then employed the symbols in a comprehensive survey. This survey quantified the impact of the four information states represented by the symbols. 


This study utilizes an experimental methodology comprising two parts: an experiment and a questionnaire. The experiment involves mentors transmitting messages to students to examine the student's comprehension of the information conveyed. Due to substantial time constraints on mentors, this experiment emphasizes enhancements in students' understanding and efficiency after receiving messages. The experiment focuses on the recipient end of the information transmission, seeking an in-depth understanding from the student's perspective regarding the influence and potential benefits of the information state symbols. By concentrating on the students' reception and processing of the transmitted information, the experiment concentrates on the recipient end of the information transmission, seeking to understand from the students' perspective the influence of the information state symbols on the students' work efficiency and level of comprehension of the mentors.


\subsection{Case One: Thesis Revision Guidance}


Paper revision is a common activity within mentor-student relationships, whether in the academic research domain or professional development. In this scenario, students often require guidance and feedback from their mentors to improve their papers. Mentors, in turn, need to carefully review the students' papers, provide comments, and suggest revisions. In this process, clear information states are crucial to enable students to understand the progress of their papers' feedback and recommendations by their mentors. Simultaneously, mentors also need to effectively communicate their review process to students, so they are aware of when to expect feedback. Therefore, this scenario underscores the evident need for symbols to denote information states, which contributes to enhancing the efficiency of information exchange and reducing unnecessary delays and misunderstandings.

To assess the effectiveness of symbols representing different informational states, we designed four corresponding scenarios for comparative experiments. These experiments aimed to investigate whether symbols play a role in eliminating uncertainties and misunderstandings, improving work efficiency, fostering academic confidence, and strengthening the bond of trust.
\begin{enumerate}
    \item[\textbf{Q1:}]  The mentor has seen my message. 
    
    The purpose of this inquiry is to examine whether symbols can convey signals of information closure, alleviating students' concerns about whether the mentor has received the message and whether there is a need to resend it.

    \item[\textbf{Q2:}] The mentor is still working on revising my paper. 
    
    The objective of this question is to evaluate whether symbols can transmit signals of task progress, informing students that the paper revision has been prioritized by the mentor, allowing them to better plan their other tasks and enhance work efficiency.

    \item[\textbf{Q3:}] The mentor is satisfied with my paper. 
    
    This inquiry aims to assess whether symbols can communicate signals of the mentor's evaluation, enabling students to realize that the mentor is willing to invest time in verifying the paper's content, thereby making the paper more accurate and refined and bolstering the student's academic confidence.

    \item[\textbf{Q4:}] The mentor is dedicated to guiding my paper. 
    
    The purpose of this question is to investigate whether symbols can convey signals of mentor support, helping students understand that the current round of paper revision is nearing completion, and the mentor will provide detailed and effective feedback, thus fostering trust between the student and mentor.

\end{enumerate}







\begin{figure*}[htb]
  \centering
  \begin{tabular}{@{}c@{\hspace{4mm}}c@{}}
    \includegraphics[width=53mm,height=36mm]{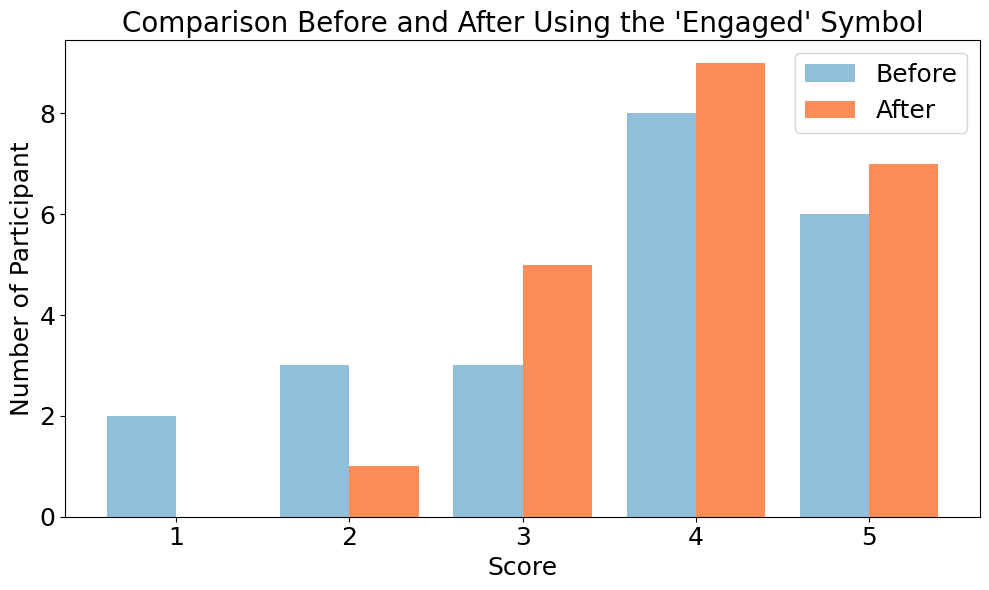} &
    \includegraphics[width=53mm, height=36mm]{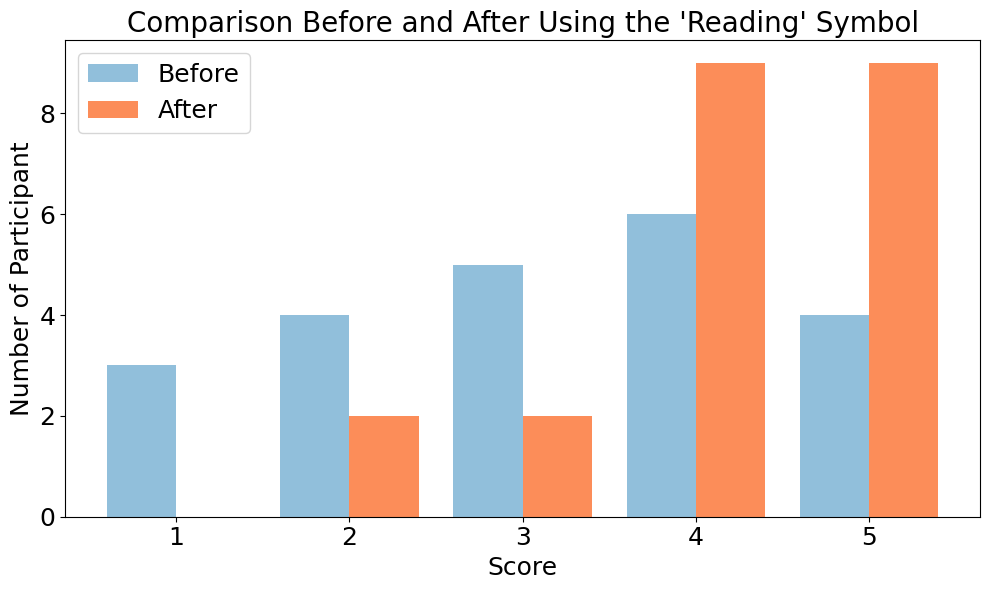} \\
    \small (a)  Comparison before and after using the "Engaged" symbol &
    \small (b)  Comparison before and after using the "Reading" symbol \\
    
    \includegraphics[width=53mm,height=36mm]{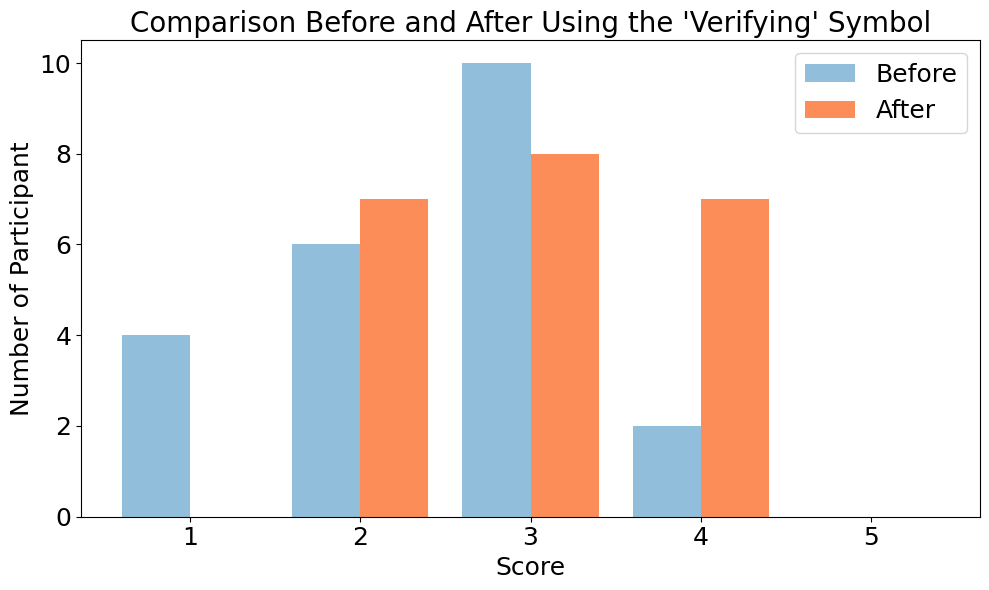} & 
    \includegraphics[width=53mm, height=36mm]{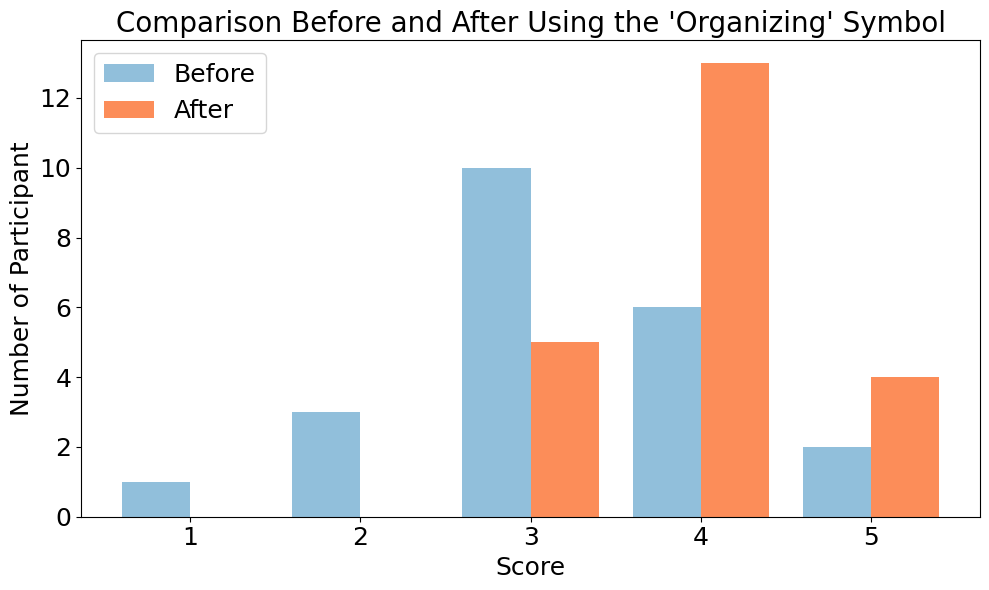} \\
    \small (c)  Comparison before and after using the "Verifying" symbol &
    \small (d)  Comparison before and after using the "Organizing" symbol  \\
    
  \end{tabular}
  \caption{Comparison of the effects of symbols used in communication between mentors and students under thesis revision guidance scenario. Agree from low to high 1-5 as strongly disagree, somewhat disagree, average, somewhat agree, and strongly agree.}
  \label{1-experiment result}
\end{figure*}

To address the four questions, we observed participants in their communication with mentors when seeing these four symbols. Subsequently, we collected participants' responses through questionnaires and recorded their reactions. The experimental results are presented in Fig. \ref{1-experiment result}. In the figures, the blue segments represent students' understanding of the tasks before using the symbols, while the orange segments represent students' understanding of mentors' tasks after using the information state symbols we designed.

It is evident that before and after using the symbols, students had a clearer understanding and more accurate awareness of the tasks their mentors were about to undertake and the actions they might expect students to follow. In Fig. \ref{1-experiment result} (a), using the "Engaged" symbol, students' average rating of their understanding of mentors' tasks was 3.59 before using the symbol and increased to 4.0 after using it, marking an 11\% improvement. In Fig. \ref{1-experiment result} (b), using the "Reading" symbol, students' average rating of their understanding of mentors' tasks was 3.18 before using the symbol and increased to 4.14 after using it, indicating a 30\% improvement. In Fig. \ref{1-experiment result} (c), using the "Verifying" symbol, students' average rating of their understanding of mentors' tasks was 2.45 before using the symbol and increased to 3.0 after using it, showing a 22\% improvement. In Fig. \ref{1-experiment result} (d), using the "Organizing" symbol, students' average rating of their understanding of mentors' tasks was 3.23 before using the symbol and increased to 3.95 after using it, marking a 22\% improvement.

In summary, for the scenario of paper revision, we observed an overall improvement in students' approval of mentors' guidance. The average approval rating increased from 3.11 to 3.77, with an average improvement of 21\% across all four questions.


\subsection{Case Two: Literature Review Guidance}

Literature review is another common activity in mentor-student relationships, particularly within the realm of academic research. Students may require guidance from mentors to select appropriate literature, comprehend key concepts within the literature, or assess the quality and relevance of academic papers. In this context, clarity regarding information states becomes crucial for students to understand whether mentors have initiated the review of literature or if they need additional time to read and comprehend the materials. Likewise, mentors need to convey their progress in information processing to inform students when their guidance is available. Therefore, this scenario also serves as an application context for the information state design, contributing to enhanced efficiency and quality of information exchange.

For the literature review guidance scenario, we have designed the following four questions to examine the functional utility of symbols.
\begin{enumerate}
    \item[\textbf{Q1:}] The mentor has seen my message. 
    
    The purpose of this query is to assess whether symbols can convey signals of information closure, thereby alleviating students' uncertainties regarding the need to resend messages and helping students anticipate the mentor's response.

    \item[\textbf{Q2:}] The mentor is reading the work. 
    
    The objective of this question is to evaluate whether symbols can communicate signals of task progress, informing students that the mentor has initiated literature review work and will provide effective answers to the queries raised by the students.

    \item[\textbf{Q3:}] The mentor considers this literature worthy of investigation.
    
    This question aims to test whether symbols can convey the mentor's attitude, allowing students to recognize the mentor's interest in the literature topic, acknowledgment of its academic value, and confidence in the student's ability to gather and organize literature, thereby assisting students in building academic confidence.

    \item[\textbf{Q4:}] The mentor will provide effective guidance. 
    
    The purpose of this question is to assess whether symbols can convey signals of mentor support, helping students understand that the mentor is willing to invest time in organizing their thoughts, providing effective guidance to resolve confusion, and offering academic and emotional support to the students.
\end{enumerate}







\begin{figure*}[htb]
  \centering
  \begin{tabular}{@{}c@{\hspace{4mm}}c@{}}
    \includegraphics[width=53mm,height=36mm]{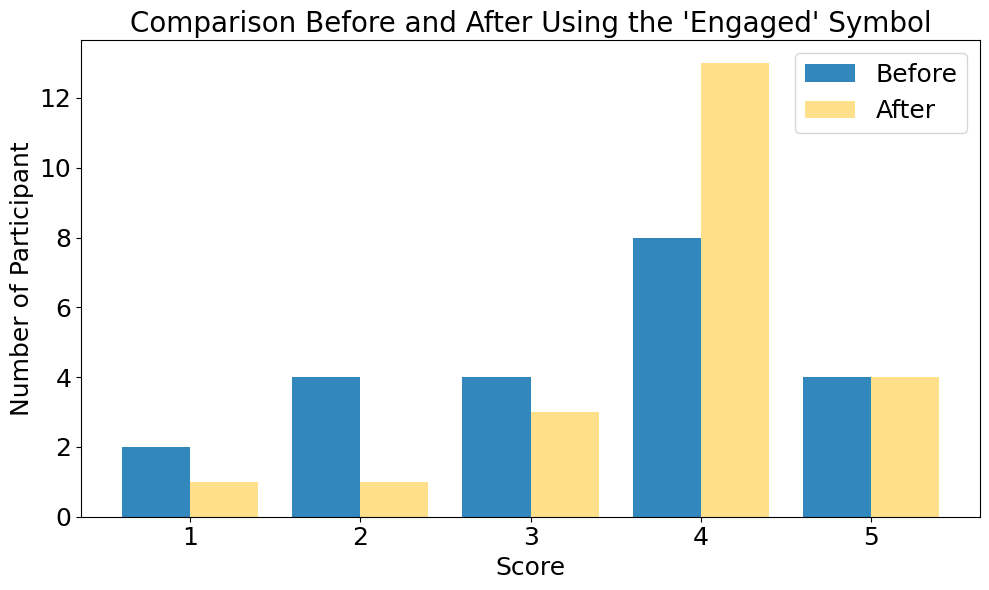} &
    \includegraphics[width=53mm, height=36mm]{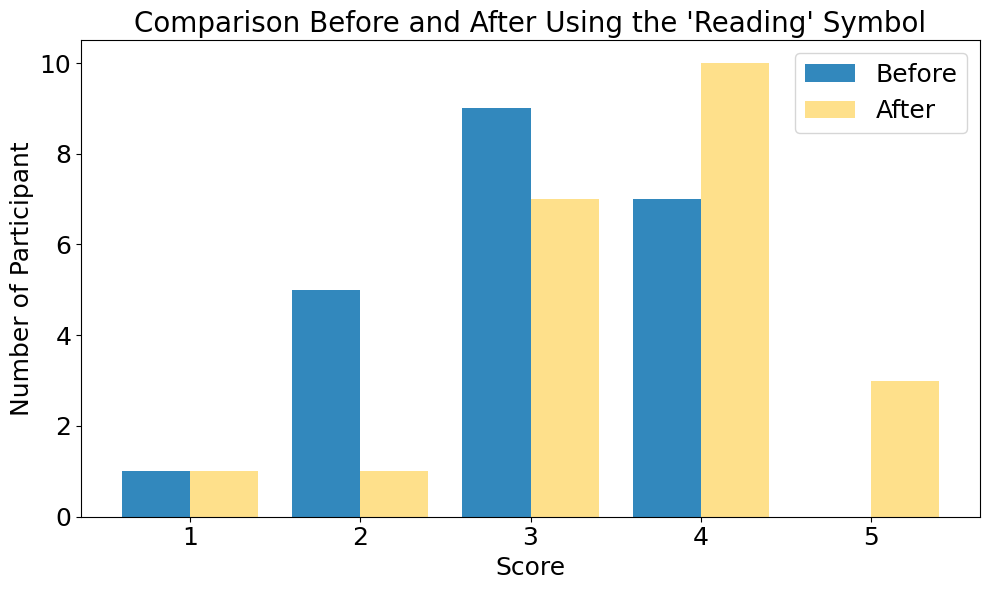} \\
    \small (a)  Comparison before and after using the "Engaged" symbol &
    \small (b)  Comparison before and after using the "Reading" symbol \\
    
    \includegraphics[width=53mm,height=36mm]{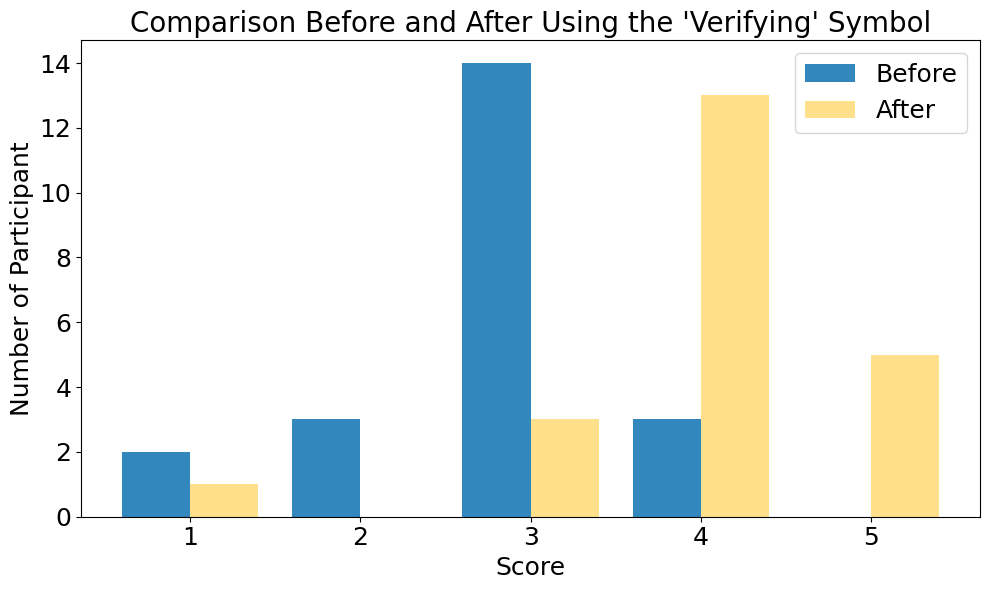} & 
    \includegraphics[width=53mm, height=36mm]{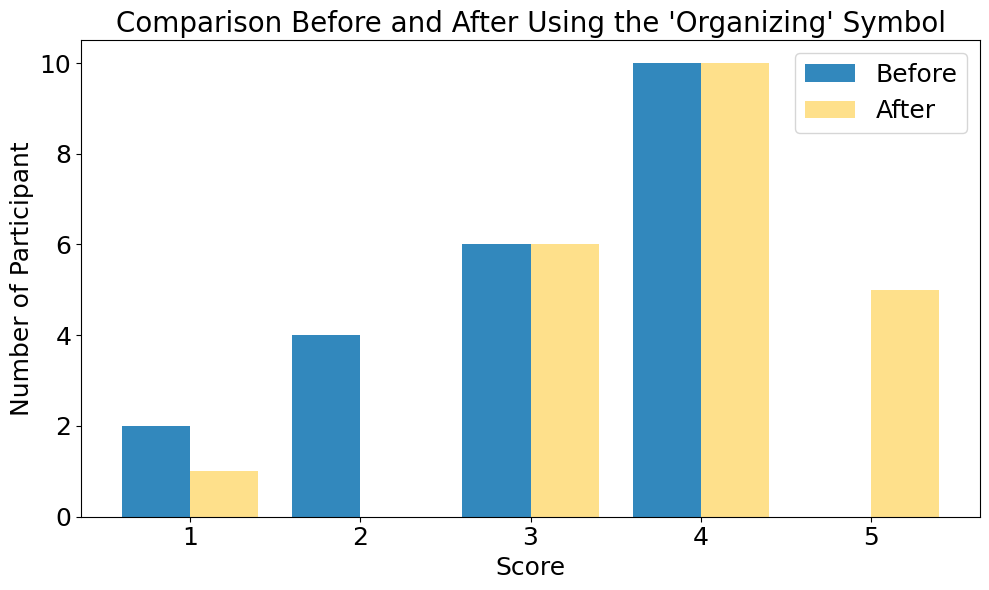} \\
    \small (c)  Comparison before and after using the "Verifying" symbol &
    \small (d)  Comparison before and after using the "Organizing" symbol  \\
    
  \end{tabular}
  \caption{Comparison of the effects of symbols used in communication between mentors and students under literature review guidance scenario. Agree from low to high 1-5 as strongly disagree, somewhat disagree, average, somewhat agree, and strongly agree.}
  \label{2-experiment result}
\end{figure*}

In addressing the four questions, we conducted observations where participants were in communication with mentors seeing these four symbols. Subsequently, we gathered feedback from participants through questionnaires and recorded their responses. The experimental results are presented in Fig. \ref{2-experiment result}. The blue segments in the figures represent students' understanding of the tasks before using the symbols, while the yellow segments indicate students' understanding of the tasks after using the information state symbols we designed. It is evident that before and after using the symbols, students had a clearer understanding and awareness of the steps their mentors were about to take and the actions they might be expected to follow.

In Fig. \ref{2-experiment result} (a), using the "Engaged" symbol, students' average rating of their understanding of the mentors' tasks before the symbol's usage was 3.36. After using the symbol, the rating increased to 3.82, marking a 13\% improvement. In Fig. \ref{2-experiment result} (b), using the "Reading" symbol, students' average rating of their understanding of mentors' tasks before the symbol's usage was 3.0, while it increased to 3.59 after symbol usage, showing an approximately 20\% improvement. In Fig. \ref{2-experiment result} (c), using the "Verifying" symbol, students' average rating of their understanding of mentors' tasks before the symbol's usage was 2.81, which increased to 3.95 after symbol usage, indicating a 40\% improvement. In Fig. \ref{2-experiment result} (d), using the "Organizing" symbol, students' average rating of their understanding of mentors' tasks before the symbol's usage was 3.09. After using the symbol, the rating increased to 3.82, showing a 23\% improvement.

Overall, for the scenario of paper revision, we observed that students' average level of approval of mentors' guidance increased from 3 to 3.8, with an average improvement of 26\% across all four questions.



\section{Discussion and Conclusion}

\subsection{Discussion}

While our study has designed a set of symbols to describe information states, this design can still be further optimized. Future research can improve these symbols through user feedback and experiments to ensure that they are more understandable and user-friendly. Additionally, exploring alternative symbols for information states to meet different contexts and needs is a viable avenue for future investigation.

The current study has primarily focused on the immediate effects of information exchange, but the long-term effects of mentorship relationships are equally deserving of attention. At this stage, the research has only examined the efficiency changes brought about by information state symbols for students after receiving messages from mentors. Further research is needed to understand the comprehension of symbols from the mentor's perspective. Future studies can delve into the long-term interactions between mentors and students and further explore mentors' understanding of these symbols to ascertain their sustained contribution to mentorship relationships. This will contribute to a deeper understanding of the practical value of information labeling.


\subsection{Conclusion}

This study introduces a more precise description of information states. In the communication between mentors and students, we have designed a set of symbols to accurately depict the cognitive status of the information being exchanged. The utilization of these symbols aids mentors and students in better labeling the status responses of information, thereby averting unnecessary misunderstandings. Through this innovation, we have addressed the issue commonly faced by students in obtaining clear instructions from mentors and the potential confusion arising from the lack of mentor responses. This innovation provides a clearer framework for communication between mentors and students, to enhance communication efficiency and improve the quality and effectiveness of mentorship. This research offers a valuable new approach in the field of information exchange, poised to have a broad impact on future educational research and practice. Through continued efforts and in-depth exploration, we can further refine the theoretical framework of this field, enhance the efficiency and quality of mentor-student relationships, and provide better support for student learning and development.


\newpage
\bibliographystyle{eg-alpha-doi}  
\bibliography{egbibsample}      

\end{document}